\documentclass[journal=aanmf6,manuscript=article]{achemso}

\usepackage[version=3]{mhchem} 

\usepackage{xcolor}
\author{Matheus Jacobs}\email{jacobs@physik.hu-berlin.de}
\author{Jannis Krumland}
\author{Caterina Cocchi}
\altaffiliation{Institute of Physics, Carl von Ossietzky Universität Oldenburg,26129 Oldenburg, Germany}
\email{caterina.cocchi@uni-oldenburg.de}
\affiliation[Humboldt-Universität zu Berlin]
{Physics Department and IRIS Adlershof,
Humboldt-Universität zu Berlin, 12489 Berlin, Germany}

\title
   {Laser-Controlled Charge Transfer in a Two-Dimensional Organic/Inorganic Optical Coherent Nanojunction}
\keywords{Ultrafast dynamics; time-dependent density-functional theory; transition metal dichalcogenides; hybrid interfaces; charge transfer; electronic coherence.}

\begin{document}


\newpage

\begin{abstract}
Understanding the fundamental mechanisms ruling laser-induced coherent charge transfer in hybrid organic/inorganic interfaces is of paramount importance to exploit these systems in next-generation opto-electronic applications.
In a first-principles work based on real-time time-dependent density-functional theory, we investigate the ultrafast charge-carrier dynamics of a prototypical two-dimensional vertical nanojunction formed by a \ce{MoSe2} monolayer with adsorbed pyrene molecules.
The response of the system to the incident pulse, set in resonance with the frequency of the lowest-energy transition in the physisorbed moieties, is clearly nonlinear.
Under weak pulses, charge transfer occurs from the molecules to the monolayer while for intensities higher than 1000~GW/cm$^{2}$, the direction of charge transfer is reverted, with electrons being transferred from \ce{MoSe2} to pyrene. 
This finding is explained by Pauli blocking: laser-induced (de)population of (valence) conduction states saturates for intensities beyond 200~GW/cm$^{2}$.
Evidence of multi-photon absorption is also provided by our results.
A thorough analysis of electronic current density, excitation energy, and number of excited electrons supports the proposed rationale and suggests the possibility to create an inorganic/organic coherent optical nanojunction for ultrafast electronics.

\end{abstract}

\newpage
\section{Introduction}
The possibility to control the electronic structure of materials \textit{via} electronic coherence has become a realistic perspective in the last few years~\cite{zhou+21as,zhou+21am}.
Hybrid inorganic/organic interfaces, combining the superior light-harvesting abilities of carbon-conjugated molecules with the efficient carrier mobility of inorganic semiconductors~\cite{blum-koch21}, including transition-metal dichalcogenide (TMDC) monolayers, are suitable platforms to host these effects and hence to be employed as functional blocks for this kind of application~\cite{kafl+19jacs,sula+20ees,chen+20nano,ulma-quek21nl}.
Recent experiments have demonstrated that, by exciting TMDC-based hybrid interfaces with a laser pulse, it is possible to induce electron or hole transfer depending on the level alignment between the components~\cite{bett+17nl,ye+21jpcl}.
In such complex heterostructures, this finding unveils intriguing opportunities to generate laser-induced photocurrents exploiting electronic coherence.
By tuning the laser frequency, one could selectively excite a specific electronic transition and resonantly inject charge carriers to the conduction band.
Furthermore, control over the field intensity enables adjusting the amount of energy transferred to the system as well as the number of photoexcited electrons and holes, thereby reproducing the working mechanism of an optical nanojunction.
Corresponding devices would have the advantage of a substantial improvement of the current speed limits, thus paving the way towards ultrafast optoelectronics~\cite{scho+19acsph}.

The successful manipulation of material properties \textit{via} laser irradiation, which is currently accessible with state-of-the-art experiments,
 requires in-depth knowledge of the electronic structure of the interface as well as thorough understanding of the microscopic mechanisms governing the coherent response of the system to the external, ultrafast perturbation.
 Theoretical models based on the semiconductor Bloch equations have proven to be particularly effective to describe the (de)coherene processes occurring in laser-excited materials of different kinds~\cite{nguy+19jpcl,hu+20jcp,chen+21jcp}.
 Also semi-empirical atomistic models such as density-functional tight-binding have been successfully employed to describe these processes~\cite{habi+19cm,xu+19nl}.
 \textit{Ab initio} methods based on real-time time-dependent density-functional theory (RT-TDDFT) offer the additional advantage of a parameter-free description of the materials and processes involved with full control over the characteristics of the incident electric pulse (shape, duration, intensity, etc.) and its temporal evolution~\cite{wach+14prl,tanc+17natcom,jaco+20apx}. 
This type of analysis is essential to identify and rationalize the fundamental mechanisms acting on a complex material during coherent laser irradiation with important implications for the design of next-generation quantum nanodevices.

With this goal in mind, in this work, we adopt RT-TDDFT to investigate laser-induced charge-transfer in a prototypical inorganic/organic heterostructure formed by a \ce{MoSe2} monolayer decorated with physisorbed pyrene molecules. 
Upon the action of a femtosecond laser pulse of varying intensity and in resonance with the lowest-energy electronic transition in the molecule, the system acts as a coherent optical nanojunction with an interfacial electric field generated in response to the excitation.
Under weak laser irradiation, when the response of the system is linear, electronic charge is transferred from pyrene to \ce{MoSe2}.
Increasing the field intensity drives the system into the nonlinear regime, causing high-harmonic generation and a sublinear increase of the excitation energy and the number of excited electrons, which both carry signatures of multi-photon absorption.
Ultimately, when the pulse overcomes the threshold of 100~GW/cm$^{2}$, electrons flow from the TMDC to the molecules, as a consequence of laser-induced saturation of valence- and conduction-band populations which in turn trigger Pauli blocking.
These results offer a proof of principle for the design of an optical nanojunction coherently controlled by femtosecond resonant pulses of varying intensity.

\section{Results}

\begin{figure}[h!]
    \centering
    \includegraphics[width=0.5\textwidth]{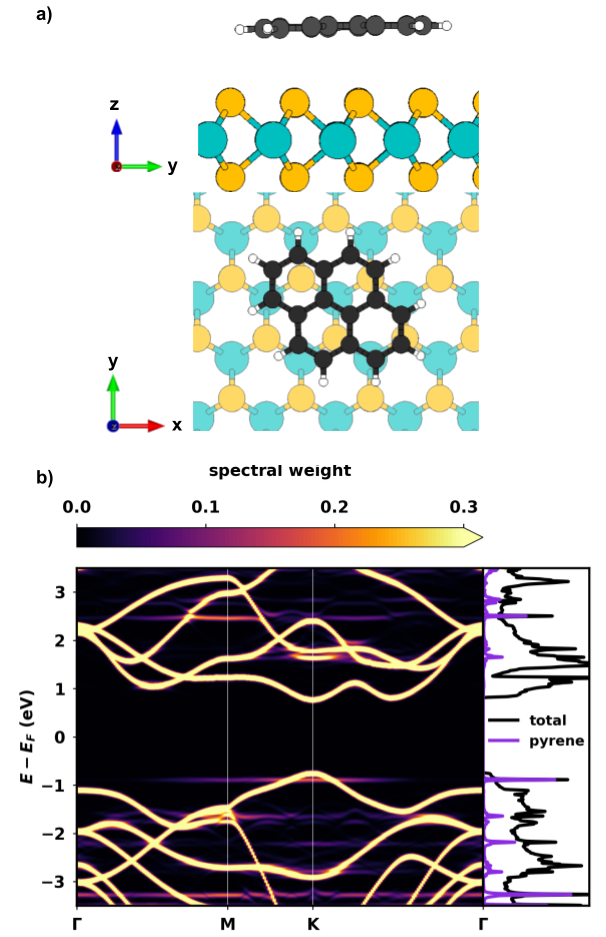}
    \caption{a) Ball-and-stick representation of the pyrene/\ce{MoSe2} interface considered in this work, where Mo atoms are depicted in turquoise, Se atoms in yellow, C atoms in grey, and H atoms in white: side view (upper panel) and top view (lower panel).
    b) Unfolded band structure of the system plotted in the unit cell of \ce{MoSe2} with the aid of the spectral function (left, color bar on top) and corresponding density of states (right) with the projected out contributions of pyrene in purple.
    }
   \label{fig.structure}
\end{figure}

The starting point of our study is the ground-state electronic structure of the two-dimensional pyrene/\ce{MoSe2} heterostructure.
We model this system by considering a pyrene molecule physisorbed on a 4$\times$4 supercell of the \ce{MoSe2} monolayer (see Figure~\ref{fig.structure}a).
In this setup, each molecule is separated by at least 5~\AA{} from its closest replica in the $(x,y)$ plane, thereby minimizing intermolecular interactions.
The band structure computed for the hybrid system is unfolded in the Brillouin zone of \ce{MoSe2} and represented with the aid of a spectral function~\cite{krum-cocc21}; in the corresponding density of states (DOS), the contributions of pyrene are projected out to better identify the energy level associated to this subsystem.
The corresponding results reported in Figure~\ref{fig.structure}b) and obtained from semi-local DFT are in agreement with higher-level results provided by the range-separated hybrid functional HSE06~\cite{hse06}, see Ref.~\citenum{krum-cocc21}. While the size of the band gap and the relative energies of the electronic levels depend on the choice of the exchange-correlation potentials, qualitative features such as the line-up of the frontier states as well the hybridization between molecular and TMDC wave-function are identically reproduced by DFT irrespective of the adopted functional.
The level alignment at the interface is of type I: the highest occupied molecular orbital (HOMO) of the molecule gives rise to a dispersionless level a few tens of meV below the valence band maximum (VBM) of \ce{MoSe2} at the high-symmetry point K.
The conduction band minimum pertains to the inorganic component as well, giving rise to a direct band gap at K for the hybrid system.
No signs of hybridization appear between the HOMO and the highest valence band of the TMDC.
On the other hand, both the HOMO-1 as well as the lowest unoccupied molecular orbital (LUMO) of pyrene are clearly hybridized with the valence and conduction bands of the TMDC around $\pm$1.5~eV, respectively.
Signatures of other occupied and virtual molecular orbitals coupling with \ce{MoSe2} bands are visible in Figure~\ref{fig.structure}b; the conditions under which these interactions occur are systematically discussed in Ref.~\citenum{krum-cocc21}.
It is worth noting that dispersion effects associated to the molecular states in the electronic structure of the interface are not related to in-plane molecule-molecule interactions but rather to electronic hybridization between the organic and inorganic components of the heterostructure.\cite{krum-cocc21}

The type-I level alignment of the pyrene/\ce{MoSe2} heterostructure, with the larger gap pertaining to the molecule (Figure~\ref{fig.structure}b), suggest favorable conditions for laser-induced charge transfer across the interface. 
Due to the strong excitonic effects dominating the absorption onset of TMDC monolayers~\cite{li+14prb}, one may anticipate photo-induced charge separation to be more efficient above the optical gap, in the region where free-carrier excitations take place~\cite{stei+17nl}.
For this reason, we impinge the hybrid system with a laser pulse of Gaussian shape, approximately 10~fs duration, and carrier frequency $\hbar \omega_0=$ 3.1~eV in resonance with the
electronic transition between the HOMO and the LUMO of pyrene (see Supporting Information, SI, Figures~S1 and S2; Table~S1).
It should be noted, however, that both molecular orbitals are in close proximity with \ce{MoSe2} bands and especially the LUMO is clearly hybridized with them~\cite{krum-cocc21}.

\begin{figure}[h!]
    \centering
    \includegraphics[width=0.5\textwidth]{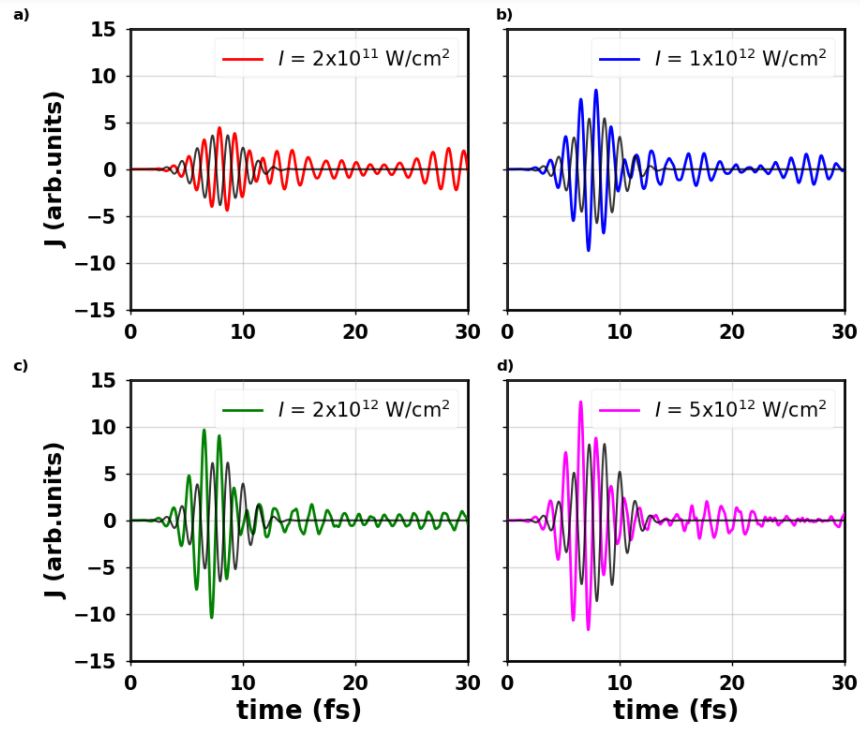}
    \caption{Time evolution of the induced electric current density, $J(t)$, at varying intensity of the laser pulse, represented in black in the background: a) $I = 2 \times 10^{11}$~W/cm${^2}$, b) $I = 10^{12}$~W/cm${^2}$, c) $I = 2 \times 10^{12}$~W/cm${^2}$, and d) $I = 5 \times 10^{12}$~W/cm${^2}$.}
    \label{fig.Current}
\end{figure}

We investigate the electronic dynamics of the hybrid system in an ultrashort time window of 30~fs in total, \textit{i.e.}, up to 20~fs after the pulse is switched off.
First, we inspect the electric current density, $J(t)$ (Eq.~\ref{eq.current}), induced by the external field to the pyrene/\ce{MoSe2} heterostructure.
With this quantity, we can monitor the response of the system to the time-dependent perturbation of varying peak intensities, ranging from $I=2\times10^{11}~\textrm{W/cm}^2$ to $I=5\times10^{12}~\textrm{W/cm}^2$ (see Figure~\ref{fig.Current}). 
During the 10~fs of laser activity, the induced current is built up. It exhibits an oscillatory behavior similar to the applied field and its amplitude varies according to the pulse intensity.
At later times, when the external perturbation is turned off, the induced current density exhibits weaker, residual oscillations, which are a signature of the remaining coherence between ground- 
and excited-states. 
Due to the essentially monochromatic nature of the pulse applied to the system, (linearly) excited states are energetically similar, resulting in coherent oscillations only within a narrow frequency window. This prevents irreversible destructive interference, instead giving rise to the observed beating pattern. In the absence of dissipation channels, these oscillations will not decay over time.\cite{bernardi+14PRL}
For the weakest considered intensities, the envelope of such oscillations and its periodicity of approximately 15~fs is visible in Figure~\ref{fig.Current}a,b.
Under the action of stronger pulses, the amplitude of these residual oscillation is considerably reduced and a periodic pattern of their envelope is hardly recognizable (Figure~\ref{fig.Current}c,d).
This behavior is associated with the non-linear response of the system~\cite{uemo+21prb}.

\begin{figure}[h!]
    \centering
    \includegraphics[width=0.5\textwidth]{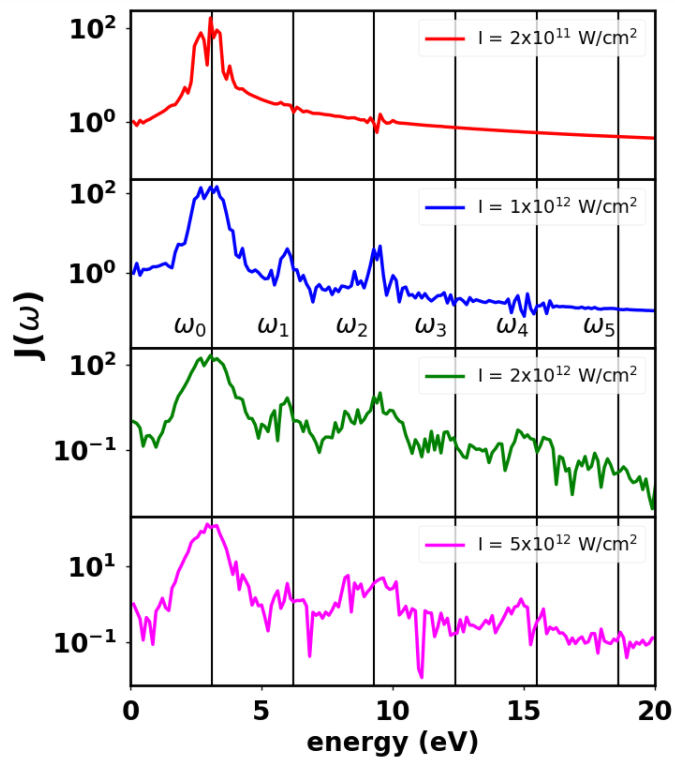}
    \caption{High-harmonic generation spectrum, $J(\omega)$,     for the pyrene/\ce{MoSe2} heterostructure impinged by pulses of increasing intensities. Vertical lines represent the fundamental carrier frequency ($\omega_0$) and its overtones ($\omega_n$ with $n=1,..., 5$).}
    \label{fig.HHG}
\end{figure}

To better clarify the nature of nonlinear processes that are in play in the considered system, we compute the harmonic spectrum, $J(\omega)$, which corresponds to Fourier transform of the time derivative of the electric current density\cite{neufeld+PRL21}  (see Eq.~\ref{eq:hhg}).
Inspection of Figure~\ref{fig.HHG} reveals that the weakest pulse excites only the fundamental harmonic corresponding to the natural carrier frequency, $\omega_{0}=3.1$~eV. 
In the spectrum obtained for $I=1\times10^{12}~\textrm{W/cm}{^2}$, maxima are present up to $\omega_{2}=9.3$~eV. 
As the intensity increases further, higher harmonics are generated up to $\omega_{4}=15.5$~eV under irradiation with $I=2\times10^{12}~\textrm{W/cm}{^2}$, and up to $\omega_{6}=21.7$~eV for $I=5\times10^{12}~\textrm{W/cm}{^2}$, which falls beyond the range displayed in Figure~\ref{fig.HHG}.
The presence of side peaks in the vicinity of the harmonic resonances is a hint of multi-photon absorption, which will be further confirmed in the remaining of our analysis below. Overall, the appearance of higher harmonics confirms the nonlinear regime of interaction between the system and the laser pulses with intensities above $10^{12}~\textrm{W/cm}{^2}$.
Analogous behavior was identified in bulk materials investigated with the same theoretical formalism~\cite{otob+08prb,zhan+20pla,zhang+PRB17}. 

\begin{figure}
    \centering
    \includegraphics[scale=0.5]{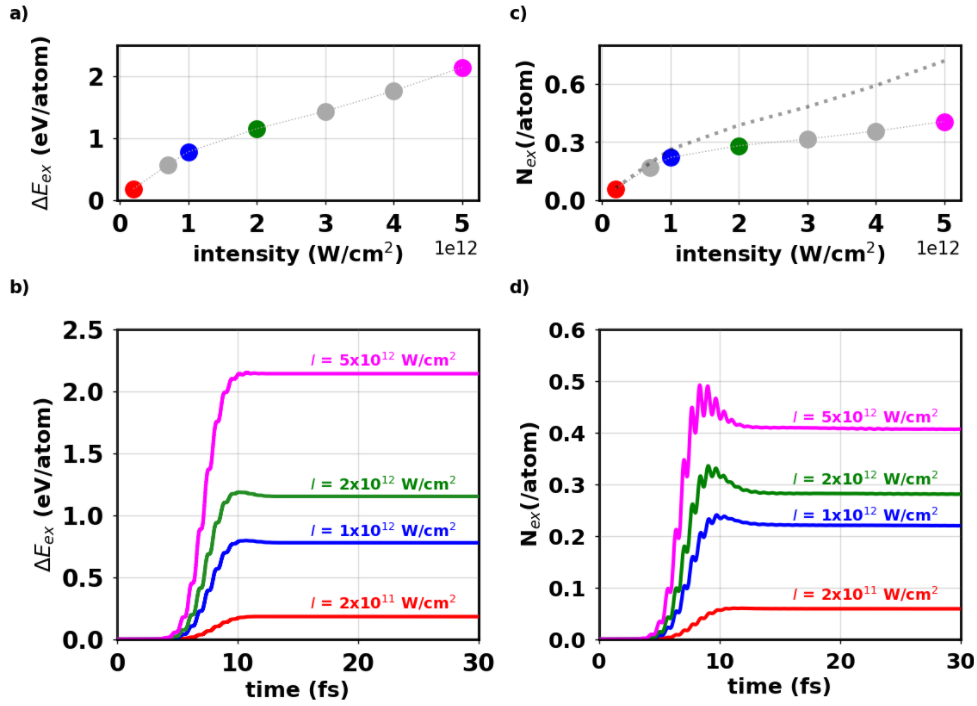}
    \caption{Excitation energy $\Delta E_{ex}(t) = E(t) - E(t=0)$ as a function of the laser intensity a) with $t=30$~fs and b) within the entire temporal range considered in these simulations. Number of excited electrons, $N_{ex}(t)$ c) with $t=30$~fs and d) within the entire temporal range considered in these simulations. In panels b) and d), the energy of the incident field is displayed in the background by a solid black line. In panel c), the dotted line indicates the hypothetical behavior of the system if the linear regime were maintained at increasing laser intensities.}
    \label{fig.elec_work}
\end{figure}

In the next step of our analysis, we monitor the excitation energy, $\Delta E_{ex}(t) = E(t) - E(t=0)$, calculated as the difference between the electronic energy at a certain time $t>0$ and in the ground state~\cite{sommer+nat16,s.sato+jcp2015}.
In Figure~\ref{fig.elec_work}a, we display this quantity calculated at $t=30$~fs for increasing pulse strengths.
The time evolution for selected intensities is shown in Figure~\ref{fig.elec_work}b following the respective color code.
Corresponding graphs referred to the gray dots in Figure~\ref{fig.elec_work}a are reported in the SI, Figure~S3.
The relation between the excitation energy and the laser intensity suggests linear response for $I=2\times10^{11} \; \textrm{W/cm}{^2}$, where $\Delta E_{ex}(t=30\,\textrm{fs})=~0.18$~eV/atom, and for $I=10^{12}\; \textrm{W/cm}{^2}$, where this value increases up to 0.77~eV/atom (see Figure~\ref{fig.elec_work}a). 
A deviation from linearity is seen starting from $I=2\times10^{12}\; \textrm{W/cm}{^2}$: in this case, the transferred energy in 30~fs increases only to 1.15~eV/atom. 
By further enhancing the laser intensity, the excitation energy keeps growing almost linearly but with a reduced slope compared to the one corresponding to the weaker pulses, reaching the value of 2.18~eV/atom at the end of the simulation window when $I=5\times$10$^{12}$~W/cm${^2}$.
The time evolution of $\Delta E_{ex}(t)$ (Figure~\ref{fig.elec_work}b) shows that after a short transient window of approximately 10~fs, coinciding with the building-up of the pulse (black line in the background) in which this quantity grows rapidly, it forms a plateau, where it remains indefinitely in the absence of dissipation effects.
The maximum value reached by $\Delta E_{ex}(t)$ during its evolution shows an intensity-dependent profile, consistent with the trends shown in Figure~\ref{fig.elec_work}a.
A similar behavior for $\Delta E_{ex}(t)$ was seen also in conventional bulk semiconductors and insulators, investigated at the same level of theory~\cite{zhan+20pla}.

Additional information about the response of the hybrid system to the applied laser perturbation can be collected from the number of excited electrons [$N_{ex}(t)$, see Eq.~\eqref{eq.exc-el}] within the considered 30~fs temporal range.
$N_{ex}(t=30~\textrm{fs})$ exhibits a more pronounced nonlinear dependence with respect to the laser intensity compared to the excitation energy computed at the end of the simulation window.
To guide the readers, in Figure~\ref{fig.elec_work}c, a dotted line marks the expected number of excited electrons in the hypothetical scenario in which the entire process occured in the linear regime.
In such a case, $N_{ex}$ = $\Delta E_{ex}$/$\hbar\omega_0$, where the carrier frequency $\omega_0$ equals the average photon frequency.
In the considered pyrene/\ce{MoSe2} interface, $N_{ex}$ follows this linear behavior up to $I = 1 \times 10^{12}$~W/cm${^2}$, in agreement with the trend exhibited by $\Delta E_{ex}$ (Figure~\ref{fig.elec_work}a).
For intensities equal or higher than $I = 2 \times 10^{12}$~W/cm${^2}$, the system is driven out of the linear regime: for $I = 5 \times 10^{12}$~W/cm${^2}$, the number of excited electrons per atom is almost halved with respect to the hypothetical linear-response scenario (Figure~\ref{fig.elec_work}c).
The discrepancy indicates an increasingly important role of multi-photon absorption for higher intensities, as the linear absorption is quenched due to saturation. This is also consistent with the rise of side peaks in the current spectrum (Figure~\ref{fig.HHG}).
The behavior of $N_{ex}(t)$ plotted in Figure~\ref{fig.elec_work}d is similar the one of $\Delta E_{ex} (t)$ in Figure~\ref{fig.elec_work}b: the value of the observable increases steeply with the ramping up of the pulse and reaches its maximum around 8~fs, after which the plateau is formed.

\begin{figure}
    \centering
    \includegraphics[width=0.9\textwidth]{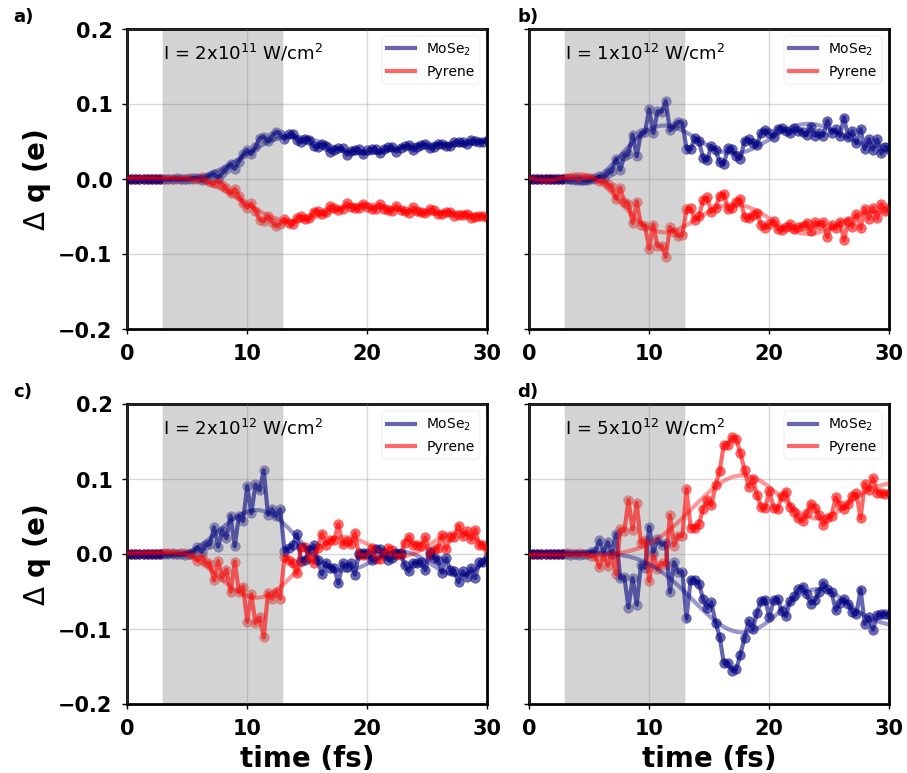}
    \caption{Time-dependent variation of the Bader partial charges with respect to the ground-state value ($\Delta q$) at intensities a) $I = 2 \times 10^{11}$~W/cm${^2}$, b) $I = 10^{12}$~W/cm${^2}$, c) $I = 2 \times 10^{12}$~W/cm${^2}$, and d) $I = 5 \times 10^{12}$~W/cm${^2}$. In the analysis, the hybrid interface is partitioned into its inorganic (MoSe$_2$) and organic component (pyrene). Positive (negative) values indicate electron accumulation (depletion). Gray areas represent the time-window in which the laser is active. The fainted lines fit of the raw data (colored dots) as a guide for the eyes.}
    \label{fig.Bader}
\end{figure}

With the knowledge gained so far from the analysis of the electric current density, of the excitation energy, and of the number of excited electrons, we are equipped to inspect the charge-transfer dynamics at the hybrid pyrene/\ce{MoSe2} interface.
These results are obtained from a time-dependent Bader charge analysis~\cite{bader90} performed throughout the considered 30~fs time window of laser-induced dynamics. 
The hybrid system is partitioned into its inorganic (\ce{MoSe2}) and organic (pyrene) components, and the partial charges are offset with respect to their ground-state value: as a result of the electronic interactions between the molecule and the monolayer, we find a slight electron excess of 0.01~$e$ on \ce{MoSe2}.
This value is of the same order of magnitude as the ground-state charge-transfer computed for molecules with recognized abilities as donors or acceptors adsorbed on a TMDC monolayer~\cite{jing+14jmca,cai+16cm,habi+20ats}.
However, it is 10 times smaller than the amount of charge withdrawn by a strong organic acceptor from a silicon surface~\cite{jaco+20apx,wang+19aelm}.

When the system is excited by the weakest considered pulse with $I = 2 \times 10^{11}$~W/cm${^2}$, a charge transfer of 0.06~$e$ occurs from the molecule to the TMDC (see Figure~\ref{fig.Bader}a). 
While this finding appears to be consistent with the type-I level alignment exhibited by the interface (Figure~\ref{fig.structure}b), we emphasize that in our simulations charge transfer takes place uniquely as an effect of laser irradiation and is not mediated by subsequent scattering events (\textit{e.g.}, due to electron-phonon coupling) redistributing electrons and holes within the bands~\cite{long+12jacs}.
With a pulse intensity of $I = 1 \times 10^{12}$~W/cm${^2}$, the charge transfer towards \ce{MoSe2} is enhanced up to a maximum of 0.10~$e$, which is reached right after the laser is switched off (see Figure~\ref{fig.Bader}b).
Overall, the time-dependent Bader charges exhibit a beating pattern resulting from the system being in a coherent superposition of multiple excited states.
This finding is consistent with the behavior of the electric current density seen in Figure~\ref{fig.Current}.
On the other hand, the fast oscillations that are visible in the raw data (blue and red dots) in Figure~\ref{fig.Bader}a,b have the frequency of the laser and are thus related to the ground-state/excited-state coherence.

When the laser intensity is doubled, reaching the value of $2 \times 10^{12}$~W/cm${^2}$, the aforementioned scenario starts to change (Figure~\ref{fig.Bader}c).   
The application of the pulse initially promotes electron transfer from the molecule to the TMDC.
However, as soon as the perturbation is turned off, the partial charges on pyrene (\ce{MoSe2}) start to change from negative to positive (positive to negative) exhibiting baseline oscillations, as a fingerprint of excited-state/excited-state coherence. 
At the end of the considered 30~fs time-window, no effective charge transfer has occurred and each subsystem is left in an essentially charge-neutral state.
The most intense of the considered lasers ($I = 5 \times 10^{12}$~W/cm${^2}$) generates a completely different response of the system with respect to the previously analyzed cases (Figure~\ref{fig.Bader}d): 0.1 electrons are transferred from the TMDC to the molecule within approximately 17~fs; subsequent oscillations appear but the excess of electrons (holes) on pyrene (\ce{MoSe2}) remains in the considered time window of 30~fs. This behavior can be additionally visualized through the charge density difference computed at the end of the simulations with respect to the ground-state value (see Figure~S4 in the SI).

\begin{figure}
    \centering
    \includegraphics[width=0.5\textwidth]{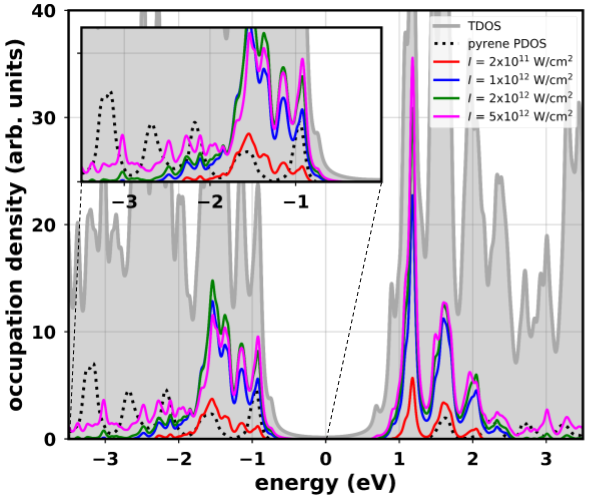}
    \caption{Occupation density (colored lines) computed at the end of the time-dependent simulations (30~fs) at increasing laser intensities. Maxima below (above) zero, where the Fermi energy is set, indicate laser-driven depopulation (population) of the electronic states. The gray area and the dotted line show the total density of states (TDOS) of the hybrid interface and the projection of the states with pyrene character (PDOS), respectively. Inset: zoom-in of the valence region. A broadening of 35~meV is applied to all displayed quantities.}
    \label{fig.occup_dens}
\end{figure}

To clarify which mechanisms drive the intensity-dependent charge transfer across the pyrene/\ce{MoSe2} heterostructure, we analyze the time-dependent occupation density (Eq.~\ref{eq.occ-density}) at the end of the 30~fs time window considered in our simulations (see Figure~\ref{fig.occup_dens}).
This quantity can be qualitatively contrasted against the total DOS of the hybrid interface and the projected contributions of pyrene (see also Figure~\ref{fig.structure}b), in order to identify those states that are mainly involved in the excitation process.  
For visualization purposes, in Figure~\ref{fig.occup_dens}, both depopulation and occupation are indicated by positive peaks below and above the Fermi energy set to zero, respectively. 

When the laser-excited system is in the linear regime ($I = 2 \times 10^{11}$~W/cm${^2}$, red curve in Figure~\ref{fig.occup_dens}), a number of valence states, including the HOMO and, in particular, the HOMO-1 of the molecule, which is clearly hybridized with valence bands of \ce{MoSe2} (see Figure~\ref{fig.structure}b), are partly depopulated. 
Corresponding occupation of conduction states is seen above the absolute minimum, involving bands either with exclusive \ce{MoSe2} character (\textit{e.g.}, around 1.1~eV in Figure~\ref{fig.occup_dens}), as well as hybridized with pyrene orbitals, \textit{e.g.}, with the LUMO, slightly above 1.5~eV. 
These trends are enhanced under increasing pulse intensity.
When the system is hit by lasers with $I=1 \times 10^{12}$~W/cm${^2}$ and $I=2 \times 10^{12}$~W/cm${^2}$ (blue and green curves in Figure~\ref{fig.occup_dens}, respectively), the valence region is remarkably depopulated. 
In the energy region between -1~eV and -2~eV, the depopulation of both the HOMO and the HOMO-1 of pyrene together with the neighboring states with \ce{MoSe2} character is evident. 
It should be noted that the HOMO-1 exhibits clear signatures of hybridization with the \ce{MoSe2} bands (see Figure~\ref{fig.structure}b).
Moreover, incipient depopulation of the HOMO-2 of the molecule around -2.2~eV, which is instead a localized molecular state, can be seen as well.
In the conduction region, a sharp increase in the occupation density up to 2~eV is visible.
The sharp maximum at $\sim$1.2~eV, corresponding to pure \ce{MoSe2} states, keep increasing from $I=1 \times 10^{12}$~W/cm${^2}$ to $I=2 \times 10^{12}$~W/cm${^2}$.
Conversely, higher-energy peaks are centered at 1.6~eV and at 2~eV and appear to be saturated for $I=2 \times 10^{12}$~W/cm${^2}$.
At the former energy we find the LUMO of pyrene, which is largely hybridized with the conduction bands of \ce{MoSe2}. 
Around 2~eV, we see states of almost exclusive \ce{MoSe2} nature (see also Figure~\ref{fig.structure}b).
Finally, under the action of the most intense pulse ($I=5 \times 10^{12}$~W/cm${^2}$, magenta curves in Figure~\ref{fig.occup_dens}), a broader energy-window of valence and conduction states participate in the excitation, enabled through multi-photon and excited-state absorption.
While the maxima between -1.8~eV and -1.0~eV in the occupation density slightly decrease in magnitude compared to their counterparts seen with weaker pulses, as a sign of achieved saturation, valence states between -2~eV and -3.5~eV start being depopulated.
Notably, in this region, a number of pyrene orbitals appear.
In the conduction band, the occupation of the \ce{MoSe2} band around 1.1~eV further increases as a function of the pulse intensity, although the small variation with respect to the result obtained for $I=2 \times 10^{12}$~W/cm${^2}$ hints toward saturation. 
On the other hand, higher-energy states including the LUMO of pyrene and the TMDC band close to 2~eV do not gain further population.
Additional unoccupied pyrene orbitals above 2.5~eV are instead involved in the excitation under the effect of this intense irradiation.

\section{Discussion}

The analysis presented so far enables us to disclose the intensity-dependent response of the pyrene/\ce{MoSe2} heterostructure excited by a femtosecond laser pulse. 
For impinging fields up to $I=1 \times 10^{12}$~W/cm${^2}$, the response of the system is linear, as testified by the relation between excitation energy and laser intensity (see Figure~\ref{fig.elec_work}); the same holds true also for the number of excited electrons per atom. 
For intensities equal or larger than $I=2 \times 10^{12}$~W/cm${^2}$, the hybrid system is driven to the nonlinear regime, as shown by the higher order harmonics in the corresponding spectrum   
(Figure~\ref{fig.HHG}).
However, the most striking intensity-dependent behavior is exhibited by the charge-transfer dynamics across the interface (Figure~\ref{fig.Bader}).
Under weak irradiation, pyrene behaves as a donor and \ce{MoSe2} as an acceptor, compatible with the type-I level alignment exhibited by this heterostructure in the ground state (see Figure~\ref{fig.structure}b).
Stronger perturbations alter this behavior, either effectively inhibiting interfacial charge transfer \textit{via} charge-density oscillations, or by reverting the direction of charge transfer, as seen for the most intense lasers considered in our simulations.

 \begin{figure}
     \centering
     \includegraphics[width=0.5\textwidth]{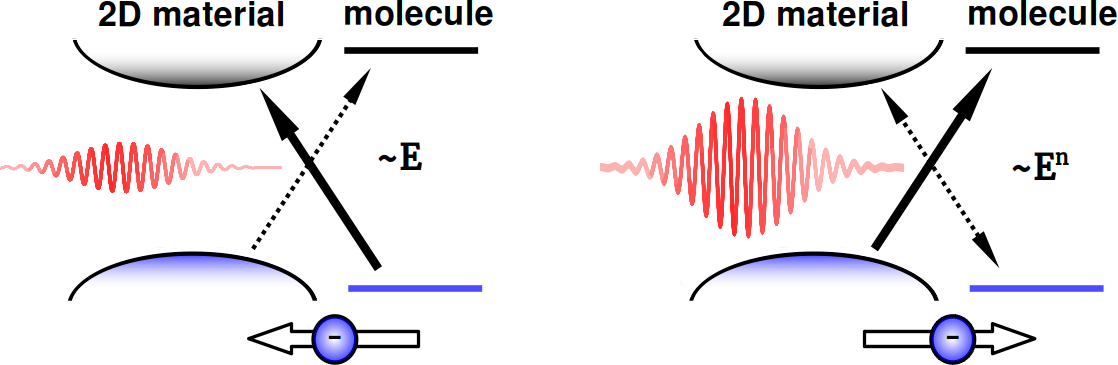}
     \caption{Pictorial sketch of the charge-transfer mechanisms occurring in the pyrene/\ce{MoSe2} heterostructure. (Left) When the system is excited by a weak pulse, the response of the system is linear with respect to the external field and the dominant mechanism is the electron transfer from the molecule to the TMDC. (Right) Upon stronger intensity driving the system in the nonlinear regime, the lowest unoccupied states of the system are saturated and the resulting direction of the charge transfer is reversed.}
     \label{fig:sketch}
 \end{figure}

The nonlinear charge-transfer behavior exhibited by the pyrene/\ce{MoSe2} heterostructure is triggered by the saturation of the lowest-unoccupied bands upon intense laser irradiation (Figure~\ref{fig.occup_dens}).
When weak pulses impinge the hybrid system, electrons are transferred from the organic to the inorganic side of the interface on account of the pulse-driven depopulation of HOMO and of the HOMO-1 of pyrene (see Figure~\ref{fig:sketch}, left panel).
While the former orbital is a localized molecular state, the latter exhibits clear signatures of hybridization with the valence bands of \ce{MoSe2}.
In turn, these transitions target both the lowest unoccupied bands of the TMDC as well as higher levels hybridized with the LUMO of the molecule (see Figure~\ref{fig.occup_dens} and Figure~\ref{fig:sketch}b).
Already under the medium pulse intensities considered in this study (I = 1000~GW/cm$^{-1}$) the aforementioned initially empty states are filled up.
As a result, pyrene$\rightarrow$\ce{MoSe2} transitions are bleached and Pauli blocking inhibits charge transfer from the molecule to the monolayer.
In contrast, \ce{MoSe2}$\rightarrow$\ce{MoSe2} and especially \ce{MoSe2}$\rightarrow$pyrene excitations take place in this scenario, leading to a depletion (accumulation) of the electronic population in the inorganic (organic) side of the interface (see Figure~\ref{fig:sketch}, right panel). 
Additional effects such as multi-photon absorption, which is triggered in the system by intense irradiation and is visible in the current spectrum and in the trend of the number of excited electrons, contribute to the nonlinear optical response of the nanojunction, too. 

\section{Conclusions and Outlook}
To summarize, in the first-principles framework of RT-TDDFT, we have investigated the laser-driven charge-carrier dynamics of a two-dimensional hybrid inorganic/organic vertical nanojunction formed by a \ce{MoSe2} monolayer and physisorbed pyrene molecules. 
By impinging the system with a coherent femtosecond pulse in resonance with the HOMO$\rightarrow$LUMO transition of the molecule and with varying intensities ranging from 200 to 5000~GW/cm${^2}$, we have found a pronounced intensity-dependent response, manifesting itself in high-harmonic generation, in the nonlinear behavior of the excitation energy and of the number of excited electrons, and, most interestingly, in the inversion of the photo-induced charge-transfer direction with respect to the linear regime.
This effect is ascribed to Pauli blocking: saturated absorption from occupied pyrene-like states to unoccupied \ce{MoSe2} bands inhibits charge transfer from the molecule to the monolayer, while promoting the same process in reverse.
Evidence of multi-photon absortion is supported by our results, too.

The presented study represents a proof of concept for a coherent optical nanojunction operated by an ultrafast resonant laser. The nonlinear response of the system to the external perturbation leads to an inversion of the charge-transfer direction according to the intensity of the applied electric field.
This behavior opens up the possibility to coherently control the sign of the interfacial electric field in the nanojunction on the femtosecond timescale, where dissipation mechanisms induced, \textit{e.g.}, by phonons are not yet relevant.
As such, our findings have important implications in the direction of ultrafast optoelectronics.

\section{Methods}

\subsection{Theoretical Background}
All calculations presented in this work are carried out the framework of RT-TDDFT~\cite{rung-gros84prl}, whereby the time-dependent Kohn-Sham (KS) equations are solved in the adiabatic approximation assuming the velocity-gauge formulation with the explicit inclusion of an electric field of the form: $\textbf{E}(t) = -\dfrac{1}{c}\dfrac{d\textbf{A}(t)}{dt}$. 
With the computed time-dependent KS orbitals, the time-dependent macroscopic current
density is evaluated as
\begin{equation}\label{eq.current}
    \textbf{J}(t) = \int_{\Omega} \text d^{3}r~\frac{1}{2}\sum_{n\textbf{k}}  \left[\phi_{n\textbf{k}}^{*}(\textbf{r},t)\left(-i\nabla + \frac{\textbf{A}(t)}{c} \right)\phi_{n\textbf{k}}(\textbf{r},t) + c.c. \right],
\end{equation}
where the integration is performed in the unit cell with volume $\Omega$.
The high harmonic spectrum of this quantity~\cite{neufeld+PRL21},
\begin{equation} \label{eq:hhg}
    J(\omega) = \left|\int dt~\partial_{t}\textbf{J}(t)e^{-i\omega t}  \right|^{2},
\end{equation}
can be used to probe high-harmonic generation.

The number of excited electrons during the laser-induced electron dynamics of the system is calculated as
\begin{align}\label{eq.exc-el}
   N_{\text{ex}}(t) = N_{\text{tot}} - \sum_{n,n',\textbf{k}}^{\mathrm{occ}}\mathcal{P}_{n',\textbf{k}}(t)|\langle\phi_{n\textbf{k}}(t=0)|\phi_{n'\textbf{k}}(t)\rangle|^{2},
\end{align}
where $\mathcal{P}_{n',\textbf{k}}(t)$ is the population of $n'$-th state at \textbf{k} and time $t$. The occupation density is calculated by weighting the density of states with the projection of the occupied time-dependent KS states into the ground-state ones at a certain time $t$:
\begin{align}\label{eq.occ-density}
   \mathcal{N}(\epsilon,t) = \sum_{n'}^{\mathrm{occ}}\sum_{n,\textbf{k}}^{\mathrm{all}}\mathcal{P}_{n',\textbf{k}}(t)\delta(\epsilon-\epsilon^{0}_{n\textbf{k}})|\langle\phi_{n\textbf{k}}(t=0)|\phi_{n'\textbf{k}}(t)\rangle|^{2}.
\end{align}

\subsection{Computational Details}

All calculations presented in this work are performed using the real-space grid code OCTOPUS,\cite{tanc+20jcp} with the exception of the band structure, as discussed below. For the approximation of the exchange-correlation functional, we adopt the adiabatic local density approximation (ALDA)\cite{perdew-zungerPRB81}, and for treat core electrons we use the Hartwigsen-Goedecker-Hutter (HGH) LDA pseudopotentials \cite{hart+98prb}. The calculations were done using a cubic grid with a spacing of 0.2~\AA,  a 3x3x1 k-grid in the super cell approach with a Coulomb cutoff of 20~\AA ~applied in the z-direction~\cite{rozzi+06PRB}.
The time-dependent electric fields are modelled as an $x$-polarized and Gaussian-envelope pulse, centered at $t_\mu=12$ fs with standard deviation $t_\sigma=2$ fs and a carrier frequency of $\omega_0 = 3.1$ eV. Simulations for different intensities ($I$) have been carried out using the approximated enforced time-reversal symmetry propagator \cite{castro+04JCP} with a time step of 2.42~as and total propagation time of 30~fs.
For the time-dependent Bader charge analysis, the time-dependent density was printed at every 0.35~fs and the charge for the TMDC and pyrene was calculated then summit the individual charges of each atoms. The time-dependent Bader charges shown in Figure~\ref{fig.Bader} are fit with an  11$^{th}$ degree Polynomial model for Support Vector 
    Regression, using the scikit-learn package.~\cite{scikit-learn}

The calculation of the band structure shown in Figure~\ref{fig.structure}b) is performed with version 6.7 of the Quantum Espresso suite\cite{qe2009, qe2020}, using norm-conserving and scalar-relativistic PZ-LDA pseudopotentials generated with the Ape code\cite{ape}. We use plane-wave cutoffs of 40 Ry and 160 Ry for the wavefunctions and the density, respectively. The density of states is calculated on a 100$\times$100$\times$1 \textbf{k}-grid by Wannier interpolation, using the Wannier90 code\cite{wannier90}. A broadening of 25~meV is applied. The contributions from pyrene are determined by projection onto the molecule-centered Wannier functions.


\begin{acknowledgement}
We thank Dieter Neher and Christian Schneider for stimulating discussions and Michele Guerrini for valuable feedback on the unpublished manuscript.
This work was funded by the German Research Foundation (DFG), project number 182087777 -- CRC 951. Additional financial support is acknowledged by C.C. to the German Federal Ministry of Education and Research (Professorinnenprogramm III), and by the State of Lower Saxony (Professorinnen für Niedersachsen).
Computational resources were provided by the North-German Supercomputing
Alliance (HLRN), project bep00104.

\end{acknowledgement}

\begin{suppinfo}
In the Supporting Information, we report the linear absorption spectra of the pyrene/\ce{MoSe2} hybrid interface as well as the analysis of the lowest excited states in gas-phase pyrene. 
We also include the time evolution of excitation energy and number of excited electrons for additional laser intensities together with the charge density difference maps at varying pulse intensity.
Refs.~\citenum{yabana+96PRB,cocc+14prl,guan+21pccp,bott+07rpp,herp+21jpca,krum+21jcp,Jamorsk+96JCP} are cited therein.

\end{suppinfo}


\providecommand{\latin}[1]{#1}
\makeatletter
\providecommand{\doi}
  {\begingroup\let\do\@makeother\dospecials
  \catcode`\{=1 \catcode`\}=2 \doi@aux}
\providecommand{\doi@aux}[1]{\endgroup\texttt{#1}}
\makeatother
\providecommand*\mcitethebibliography{\thebibliography}
\csname @ifundefined\endcsname{endmcitethebibliography}
  {\let\endmcitethebibliography\endthebibliography}{}

\newpage

\section*{TOC Graphic}
\begin{figure}[H]
    \centering
    \includegraphics{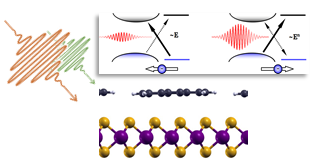}
\end{figure}

\end{document}